\documentclass{article}

\usepackage{arxiv}

\usepackage[utf8]{inputenc} % allow utf-8 input
\usepackage[T1]{fontenc}    % use 8-bit T1 fonts
\usepackage{lmodern}        % https://github.com/rstudio/rticles/issues/343
\usepackage{hyperref}       % hyperlinks
\usepackage{url}            % simple URL typesetting
\usepackage{booktabs}       % professional-quality tables
\usepackage{amsfonts}       % blackboard math symbols
\usepackage{nicefrac}       % compact symbols for 1/2, etc.
\usepackage{microtype}      % microtypography
\usepackage{graphicx}

\title{Democapsid}

\author{
    Daniel Antonio Negrón
   \\
    Bioinformatics and Computational Biology \\
    George Mason University \\
  Manassas, VA 20110, USA \\
  \texttt{\href{mailto:dnegron2@gmu.edu}{\nolinkurl{dnegron2@gmu.edu}}} \\
   \And
    Antoni Luque
   \\
    Department of Biology\\
Department of Physics\\
Department of Chemical, Environmental, \& Materials Engineering \\
    University of Miami \\
  Miami, FL 33146, USA \\
  \texttt{\href{mailto:antoni.luque@miami.edu}{\nolinkurl{antoni.luque@miami.edu}}} \\
  }

\providecommand{\tightlist}{%
  \setlength{\itemsep}{0pt}\setlength{\parskip}{0pt}}

\usepackage{hyperref}
\begin{document}
\maketitle

\begin{abstract}
Capsids are the protein shells that protect the genetic material of
viruses. The precise structural description of capsids informs how
viruses assemble and evolve and is key to the development of antiviral
targets. Most viruses form icosahedral capsids; among these, most adopt
quasi-spherical shapes, and some form elongated architectures. However,
elongated capsids have been understudied, despite their decoupling of
width and length providing greater control over their packaging
capacity, a feature of particular interest in capsid evolution and in
virus-based biotechnological platforms. A key bottleneck is the lack of
tools for the analysis and design of elongated viral capsids. To that
end, this article introduces Democapsid as a versatile tool for
generating coordinates of both quasi-spherical and elongated (and
shrunk) icosahedral capsids, as well as for producing customizable
graphical models and publication-quality figures. The underlying
algorithm builds on the generalized geometrical theory of viral capsids
and employs numerical methods to assemble capsid elements based on
folding constraints. It includes parameters controlling protein tiling
associated with the eight regular icosahedral lattices, elongation axes
(5-fold, 3-fold, and 2-fold), sphericity, and discrete body length for
prolate (extended) and oblate (shrunk) shapes. It is available as a
JavaScript
\href{https://github.com/dnanto/democapsid}{browser application}, a
\href{https://github.com/dnanto/pydemocapsid}{Python package} powering
plugins for
\href{https://github.com/dnanto/hkcage/tree/pydemocapsidize}{UCSF ChimeraX}
and \href{https://github.com/dnanto/blendocapsid}{Blender}, and an
\href{https://github.com/dnanto/rdemocapsid}{R package} for generating
reproducible documents with embedded models. The code (MIT License) is
available on GitHub. Democapsid will benefit both researchers and
graphic designers by enabling the investigation and communication of
research on viral capsids and other icosahedral compartments.
\end{abstract}

\keywords{
    virus
   \and
    capsid
   \and
    icosahedral
   \and
    elongated
   \and
    prolate
   \and
    oblate
   \and
    ChimeraX
   \and
    Blender
   \and
    python
   \and
    R
   \and
    JavaScript
  }

\section{Introduction}\label{introduction}

Capsids are protein shells assembled from multiple copies of one or more
proteins encoded in the genome of viruses. They provide essential
functions through the virus replication cycle, including genome
packaging and shielding, protrusions for cellular entry and exit, and,
for some, the assembly of an envelope layer \(\cite{lucas2002}\). The
capsid surface and its protein-protein interfaces offer reliable targets
for antivirals \(\cite{schlicksup2020}\), especially because they are
more evolutionarily conserved than other molecular targets, such as the
polymerase \(\cite{fu2019}\). The assembly of capsids has also been
engineered to form virus-like particles (VLPs) used for drug delivery,
gene therapy, vaccination, and other nanotechnology applications
\(\cite{chung2020,douglas2002}\).

The majority of capsids observed in nature and used in biotechnological
applications are quasi-spherical and organize their proteins following
icosahedral symmetry
\(\cite{louten2016,twarock2019,parent2026,castón2024}\). However, about
10\% to 20\% of viral capsids form elongated structures
\(\cite{ackermann2007}\). These capsids expand or shrink by
incorporating or removing proteins on the equatorial part of the
icosahedral architecture
\(\cite{fang2022,woodson2025,bárdy2020,hawkins2021,li2025,chang2022,luque2010a}\).
A major characteristic of elongated capsids is that their length can be
modified following small discrete steps providing a more precise
variation of the internal capsid volume
\(\cite{moody1999,twarock2004,luque2010b}\). Instead, changes in their
radius, as for quasi-spherical icosahedral capsids, are associated with
changes in the icosahedral T-number, yielding a larger change in
internal volume \(\cite{luque2020}\). The tight control in the length of
elongated capsids provides a mechanism to investigate the evolution of
changes in viral genome length and the possibility to have capsid-based
technologies with more control on varying the cargo capacity. However,
despite these structural advantages, elongated capsids have been
understudied. A bottleneck in the field is the lack of accessible tools
to model and characterize elongated capsids.

Previous tools that offer the possibility of designing elongated
structures include FULLERENE \(\cite{schwerdtfeger2013}\) and Nanotube
Modeler \(\cite{johnson1997,melchor2004}\). These two packages were
developed in the context of carbon nanotubes and other chemical-inspired
structures. They can produce a wide range of chemical structures, which
share some parallel elements to viral capsids. However, their
parameterization does not accommodate the standard properties that are
used in the reconstruction and characterization of viral capsids,
including the triangulation number (T-number), elongation number
(Q-number), axis symmetry, or sphericity
\(\cite{luque2010b,caspar1962,aznar2012}\). They do not include the
lattices associated with the generalized theory of icosahedral capsids
that can accommodate the formation of different protein oligomeric tile
shapes and minor proteins \(\cite{twarock2019}\). Additionally, the
process of illustrating and working with capsid models conceptually also
lacks versatile tools. There are two popular websites and databases in
the field, ViralZone \(\cite{hulo2011}\) and VIPERdb
\(\cite{montiel-garcia2020}\), which contained pre-rendered 2D images of
capsid cartoons for different viruses in vectorial and raster images,
but they do not offer the possibility to modify the icosahedral
parameters.

The Democapsid project presented here, formerly ``capsid.js''
\(\cite{negrón2021}\), addresses these two issues mentioned above.
Within the kernel of Democapsid is a mathematical algorithm that has
been developed to generate capsid coordinates that can be modified based
on the key parameters for quasi-spherical and elongated viral capsids,
including surface icosahedral lattice parameters, hexagonal lattice
steps controlling the configuration of pentameric vertices in
quasi-spherical (T-number) and elongated/shrunken architectures (T-,
Q-numbers), longitudinal elongation axes (5-fold, 3-fold, and 2-fold),
radius (length scale), and sphericity. The rendering options provide
control over the color and transparencies of lattice elements. Computed
structures can be explored in a variety of numerical and graphical
formats, facilitating the use of the outputs as structural models or as
2D or 3D editable capsid cartoons.

All Democapsid project code is available on GitHub under the MIT
license, including the browser application
(\url{https://github.com/dnanto/democapsid}) and Python package
(\url{https://github.com/dnanto/pydemocapsid}). Plugins were developed
using the latter for the UCSF ChimeraX structural biology platform
\(\cite{meng2023}\) (\url{https://github.com/luquelab/hkcage}) and
Blender graphic design suite \(\cite{blenderfoundation2024}\)
(\url{https://github.com/dnanto/blendocapsid}). An R package
(\url{https://github.com/dnanto/rdemocapsid}) was also developed by
wrapping the underlying JavaScript of the browser application. Notably,
the JavaScript implementation provides Scalable Vector Graphics (SVG)
export. This is ideal, since it is decomposable, infinitely scalable,
and compatible with common office productivity software and graphics
programs, thereby allowing scientists, students, teachers, and artists
to produce publication-quality, theory-consistent images. Democapsid,
thus, is also applicable to other icosahedral based structures beyond
viral capsids, such as cellular microcompartments and cages
\(\cite{parent2026}\).

\section{Modeling software}\label{modeling-software}

\begin{figure}
  \centering
  \fbox{\includegraphics[width=1.0\textwidth]{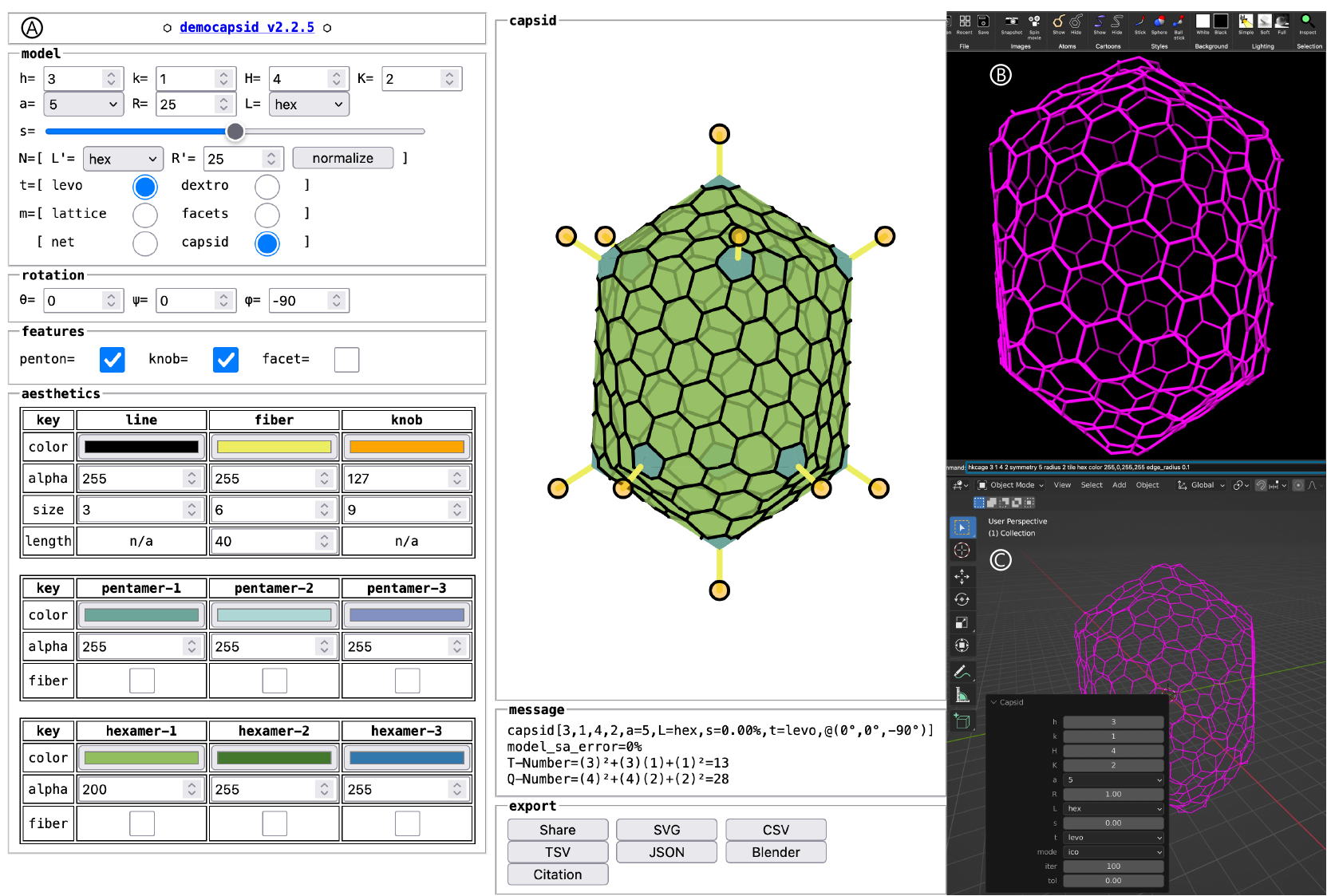}}
  \caption{Democapsid software. The Democapsid JavaScript browser app (A) provides several modules to generate models and export results. Plugins powered by the pydemocapsid Python package include the updated hkcage UCFS ChimeraX plugin (B) and the new blendocapsid add-on for Blender (C). Examples correspond to a $T(h = 3, k = 1) = 13$ and $Q(H = 4, K = 2) = 28$ architecture corresponding to the Enterobacteriophage T4 capsid. Each program exposes the same architecture parameters.}
  \label{fig:1}
\end{figure}

Figure \(\ref{fig:1}\) contains a screenshot of the Democapsid
JavaScript browser application and Python-based hkcage UCSF ChimeraX
(v1.11.1) plugin and blendocapsid Blender (\(\ge\)v3.0) add-on. The
Paper.js (v0.12.18) library provided support for SVG composition and
geometry within the democapsid app \(\cite{lehni2020}\). Plugins were
built around the pydemocapsid Python package, which uses the NumPy
(\(\geq\)v1.16.5) package for lineage algebra \(\cite{harris2020}\). The
R package was built using the htmlwidgets (v.1.6.4) and shiny (1.13.0)
frameworks \(\cite{vaidyanathan2023,chang2026}\). These tools are free
and open-source and available on GitHub under the MIT License.

\subsection{Browser application}\label{browser-application}

The JavaScript browser app is divided into seven modules (Figure
\(\ref{fig:1}\)A): \textbf{\emph{Model}}, \textbf{\emph{Rotation}},
\textbf{\emph{Features}}, \textbf{\emph{Aesthetics}},
\textbf{\emph{Capsid}}, \textbf{\emph{Message}}, and
\textbf{\emph{Export}}. Models are generated completely within the
client browser. The app is a fully self-contained HTML file and can be
run offline.

\textbf{\emph{Model module.}} Model parameters control the capsid
architecture. The walk parameters define the magnitude (\(h\), \(k\),
\(H\), and \(K\)) and turn (\(t\)) direction over the lattice (\(L\)) as
described in Figure \(\ref{fig:2}\). Together, they define the
triangular cap and body facets for transformation onto the vertices for
the selected axial symmetry \(a\) as described in Figure
\(\ref{fig:3}\). Capsid size also scales with the circumradius \(R\) of
the lattice primitive unit. The \(s\) parameter sets the percentage
interpolation of the capsid vertices to the cap circumsphere and body
cylinder. Different display modes are available via the \(m\) parameter
to render the lattice, facets, foldable net, and capsid models.
Additionally, the \(N\) parameter group normalizes the size of the
current model based on another tiling primitive \(L’\) and its
circumradius \(R’\).

\textbf{\emph{Rotation module.}} The yaw (\(\theta\)), pitch
((\(\psi\)), and roll (\(\phi\)) angles, with values in degrees, control
the orientation of the capsid by setting the rotation matrix of the
camera model.

\textbf{\emph{Features module.}} Feature controls toggle whether to
render penton vertex fibers, a fiber knob, or to draw an outline for
each triangular facet.

\textbf{\emph{Aesthetics module.}} Aesthetic controls are subdivided
into three tables. The first controls elements of the
\textbf{\emph{Features}} module, such as color, size/width,
transparency, and length. Parameters for the color and transparency of
pentamers and hexons are available in the latter two. A checkbox toggles
a fiber projection from the centroid of each subunit for the
corresponding penton or hexon.

\textbf{\emph{Capsid module.}} An HTML canvas element displays the model
using the Paper.js JavaScript library.

\textbf{\emph{Message module.}} The message box lists the parameters,
surface area error, and triangulation numbers \(T(h,k)=h^2+hk+k^2\) and
\(Q(H,K)=H^2+HK+K^2\) of the viral cap and body, respectively.
Parameters are listed within square brackets with \(h\), \(k\), \(H\),
and \(K\) as the first four numerical entries. Model error corresponds
to the deviation from the expected surface area of the capsid facets
following an affine transformation to the computed vertices, which can
result in distortion if the numerical solution is sub-optimal.

\textbf{\emph{Export module.}} The export module saves the model as a
shareable link, an SVG, a set of points and edges as a delimited table
(CSV/TSV), a structured JSON file, or a Blender script. Shareable links
include a query parameter containing the BASE64-encoded
\textbf{\emph{Model}}, \textbf{\emph{Rotation}},
\textbf{\emph{Feature}}, and \textbf{\emph{Aesthetics}} values. SVG
files represent the model as two-dimensional lines and polygons ordered
by depth and hierarchically grouped by lattice primitive and facet. The
tabular file lists the three-dimensional model points, with keys that
group them by facet, polygon, and segment. Similarly, the JSON file
groups the points as a series of nested arrays. The script contains the
coordinate data and can be run directly in the ``Scripting'' tab of the
Blender application without needing to install plugins or dependencies.

\subsection{Packages \& plugins}\label{packages-plugins}

The pydemocapsid Python package provides the same functionality as the
JavaScript version on the command-line interface (CLI), but it only
generates vertices or meshes as coordinates. It exposes the same
parameters available in the Model module and powers the hkcage plugin
for rendering structures within the UCSF ChimeraX (Figure
\(\ref{fig:1}\)B) molecular modeling software and the blendocapsid
add-on for the Blender (Figure \(\ref{fig:1}\)C) 3D graphics program.
The hkcage plugin facilitates comparative modeling and structural
analysis of capsid models to crystallographic data with other tools in
the ChimeraX ecosystem. As a wrapper for the JavaScript library, the
rdemocapsid R package can reproduce the browser application as a static
page or dynamic shiny application. For artistic objectives, such as
scientific communication, blendocapsid generates meshes for rendering 3D
models.

\section{Algorithmic innovation}\label{algorithmic-innovation}

Previous tools lacked the ability to generate elongated capsids and
vector-based, decomposable figures. The approach here starts with
established 2D-3D mapping geometric theory \(\cite{luque2010b}\) to
compute the required capsid facets and introduces two algorithmic
innovations. The first regards the generation of coordinates for
elongated capsids. Starting with invariant cap coordinates, facets are
added and rotated around key edges such that they respect constraints
with respect to cap or body circumference, which can be done with
numerical methods. The second innovation groups the resulting mesh by
lattice primitive, facet, and capsid, allowing the user to decompose the
structure in SVG-compatible software.

\subsection{Capsid coordinates}\label{capsid-coordinates}

\begin{figure}
  \centering
  \fbox{\includegraphics[width=1.0\textwidth]{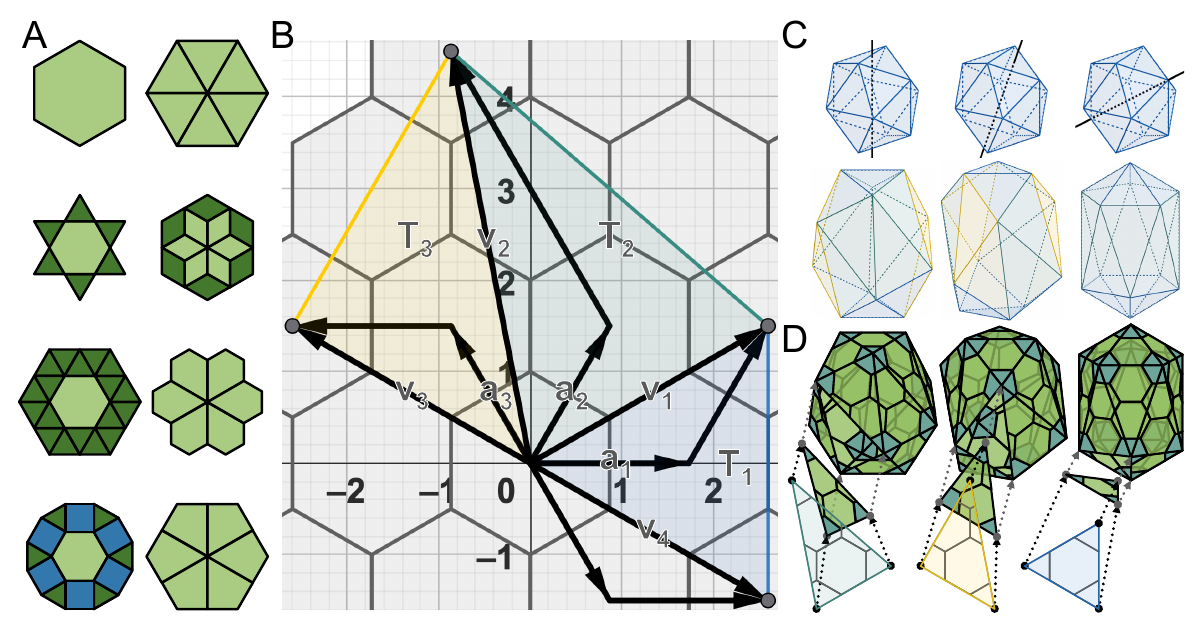}}
  \caption{Capsid construction overview. A) The selected lattice primitive is a hexagon with a circumradius $R = 1$. B) Walk parameters are $h = 1$, $k = 1$, $H = 1$, $K = 1$, and $t = levo$ (levo turn). The $\vec{a}$ vectors denote the basis walk vectors. Linear combination with the walk parameters and vectors defines the facet vectors $\vec{v}$ and the points of each triangular facet $T$. C) Construction varies according to the selected axial symmetry. The top row shows the 2-, 3-, and 5-fold symmetry options. The bottom row is the result of arranging the triangular facets according to the constraints of each symmetry. D) An affine transformation step places each triangular lattice piece according to the coordinates obtained from the previous step. This figure contains elements designed with GeoGebra \cite{hohenwarter2002}.}
  \label{fig:2}
\end{figure}

The 2D geometric theory of viral capsids was first applied to generate
the triangular facets based on the selected walk magnitude (\(h\),
\(k\), \(H\), and \(K\)), turn \(t \in \{levo, dextro\}\), and lattice
primitive tile \(L\) (having Schläfli symbol \(\{6,3\}\)). These
parameters determine the grid and vectors used to carve-out the facets
of the capsid. Accordingly, the tile circumradius (\(R\)) and basis
(\(B\)) define the grid for the walk. In the case of the hexagonal
lattice, \(B = r [ 2, 1; 0, \sqrt3 ]\), where \(r = R \sqrt3 / 2\) is
the inradius and the matrix represents the translation vectors. The
vectors \(\vec{a}_1\) and \(\vec{a}_2\) correspond to the first two row
vectors of B and \(\vec{a}_3\) is the 60° rotation of the first. Table
\(\ref{tbl:1}\) derives the vectors outlining the capsid facets.

\begin{table}
 \caption{Capsid facet vectors are based on positive, non-zero, discrete walk ($h$, $k$, $H$, and $K$) and turn $t$ parameters. The vectors $\vec{a}_1$ and $\vec{a}_2$ correspond to the basis of the selected tiling primitive and $\vec{a}_3$ is the 60° rotation of $\vec{a}_1$.}
  \centering
  \begin{tabular}{lllll}
    \toprule
    turn   & $\vec{v}_{1}$ & $\vec{v}_{2}$ & $\vec{v}_{3}$ & $\vec{v}_{4} $  \\
    \midrule
    levo   & $ha_1 + ka_2$ & $Ha_2 + Ka_3$ & $Ha_3 - ka_2$ & $-ha_3 + ka_1$ \\
    dextro & $ha_1 - ka_3$ & $Ha_2 + Ka_1$ & $Ha_3 + ka_2$ & $-ha_3 - ka_2$ \\
    \bottomrule
  \end{tabular}
  \label{tbl:1}
\end{table}

\begin{figure}
  \centering
  \fbox{\includegraphics[width=1.0\textwidth]{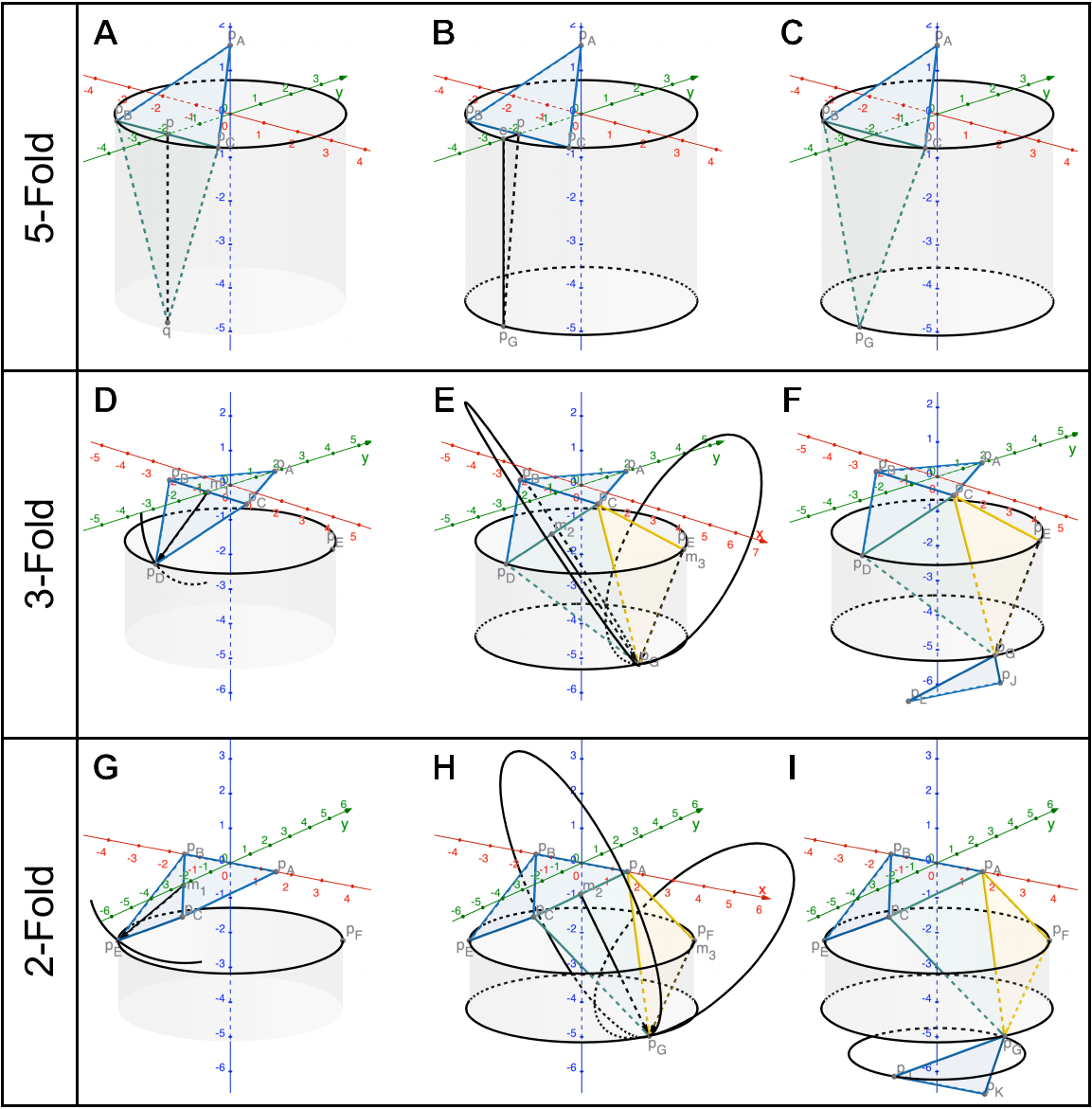}}
  \caption{Axial construction summary. All constructions begin with the cap since the points are invariant. Cylindrical body constraints determine how successive facets are added to determine the minimum number of points before symmetry can account for the rest. For the 5-fold, a right-triangle is constructed based on (A) the projection of the body facet tip (point $q$) to the bottom edge of the cap facet (point $p$) (B) and its subsequent projection to the rim (point $o$) via the quadratic equation, and then again onto the bottom rim (point $pG$), (C) forming the body facet. For the 3- and 2-fold, (D) (G) the cap rim is determined by folding the cap facet, obtaining point $pE$. (E) (H) A numerical method rotates the next body facet around the edge of the cap facet until it intersects the body cylinder, and the neighboring body facet also fits. (F) For 3-fold, the bottom cap facet points are obtained by symmetry. (I) These are solved numerically in the 2-fold case by rotating a point on the bottom rim until and equilateral triangle occurs in the expected position. The rest of the points are obtained by symmetry. Subfigures were generated with GeoGebra \cite{hohenwarter2002}}
  \label{fig:3}
\end{figure}

The theory and approach discussed so far provides an exhaustive
description of the discrete structures that can be mapped from 3D
structures onto a 2D map but does not explain how to generate the
coordinates of the structures algorithmically. In this work, coordinates
for the regular/five-fold symmetry cases were solved analytically
whereas the 3- and 2-fold cases required numerical methods. The latter
employed a combined bracketing and bisection algorithm to find the angle
within a range that satisfied the folding constraints. In all cases, the
procedure calculates the topmost points of the corresponding cap and
then solves the intersection of the next one or two body facets with the
symmetry-preserving body cylinder starting at the cap rim. The regular,
five-fold, and two/three-fold cases require up to one, two, and three
different body facets, respectively. This was approached differently for
each axis of symmetry, 5-fold, 3-fold, and 2-fold (Figure
\(\ref{fig:3}\)).

\subsection{Hierarchical elements}\label{hierarchical-elements}

To facilitate post-production of figures using capsid architectures, SVG
elements were organized hierarchically so they could be accessed as
independent objects (Figure \(\ref{fig:2}\)C and \(\ref{fig:2}\)D). This
was achieved by grouping polygons within the Paper.js library. It also
allowed for direct extraction of facets from the grid via intersection
with each triangular facet. The resulting model can be imported into any
software that supports vector images, including office productivity
software. Such files can then be easily decomposed, remixed, and
annotated.

\subsection{Sphericity/Cylindricity}\label{sphericitycylindricity}

To approximate rounder viral particles, a spherization operation
interpolates surface coordinates to the icosahedron circumsphere. For
variable body length capsids, this procedure interpolates surface
coordinates to the constraining cylinder. Separate functions calculate
the spherical caps based on axial symmetry. For five- and three-fold
symmetry, the interpolation sphere is placed at the center of the circle
passing through the topmost cap point and any of the ones at the joining
body rim. In the two-fold case, the interpolation sphere is placed at
the center of the circumsphere enclosing the two top-most cap points and
the lower two at the rim.

\section{Results}\label{results}

The Democapsid project developed software that can generate accurate
models across the various lattice and symmetry combinations observed in
naturally occurring icosahedral viral capsids. It is also the first
program to realize models for 3- and 2-fold axis constructions using a
numerical method approach. This project also yielded a GPU shader that
employs signed-distance fields to generate capsids with regular symmetry
(\url{https://www.shadertoy.com/view/dlsGRH}). The goal is to then
extend this solution to 2- and 3-folds to provide a faster raster-based
preview of the SVG models. Additionally, the shader solution avoids
artifacts in the current JavaScript version since it applies the simple
painter's algorithm to orders SVG objects and emulate a 3D effect. An
option to switch to an alternative model, such as the nanotube body
approximation (20), is also possible, and would benefit the
nanotechnology field.

\section*{Conflict of Interest}\label{conflict-of-interest}
\addcontentsline{toc}{section}{Conflict of Interest}

None declared.

\section*{Author Contributions}\label{author-contributions}
\addcontentsline{toc}{section}{Author Contributions}

Conceptualization: DN \& AL. Data curation: DN. Formal analysis: DN.
Funding acquisition: AL. Investigation: DN. Methodology: DN \& AL.
Project administration: DN \& AL. Resources: DN \& AL. Software: DN.
Supervision: AL. Validation: DN. Visualization: DN. Writing -- original
draft: DN. Writing -- review \& editing: AL.

\section*{Acknowledgements}\label{acknowledgements}
\addcontentsline{toc}{section}{Acknowledgements}

A.L. acknowledges support from the National Science Foundation (Award
\#2424579). D.N. acknowledges support from the Demoscene community.

\section*{Supplementary Material}\label{supplementary-material}
\addcontentsline{toc}{section}{Supplementary Material}

\begin{itemize}
\tightlist
\item
  democapsid: \url{https://github.com/dnanto/democapsid}
\item
  pydemocapsid: \url{https://github.com/dnanto/pydemocapsid}
\item
  rdemocapsid: \url{https://github.com/dnanto/rdemocapsid}
\item
  blendocapsid: \url{https://github.com/dnanto/blendocapsid}
\item
  hkcage: \url{https://github.com/dnanto/hkcage/tree/pydemocapsidize}
\item
  protocapsid: \url{https://www.shadertoy.com/view/dlsGRH}
\end{itemize}

\bibliographystyle{unsrt}
\bibliography{references.bib}

\end{document}